\documentstyle{article}

\newcounter{eqnletter}[equation]
\catcode`\@=11
\@addtoreset{equation}{section}   
\catcode`\@=12

\newcommand{\intcauchy}{\mskip 3mu -\mskip-18mu \int}

\begin{document}

\begin{center}

{\LARGE\bf Quartic Anharmonic Oscillator  \\[5mm]
And Random Matrix Theory}

\vskip 1cm

{\large {\bf G.M. Cicuta} }

\vskip 0.1 cm

Dipartimento di Fisica, Universita di Parma,\\
Viale delle Scienze, I - 43100 Parma, Italy \footnote{E-mail
address: cicuta@parma.infn.it} \\[0.5cm]

{\large {\bf S.Stramaglia} }

\vskip 0.1 cm

Dipartimento di Fisica, Universita di Bari,\\
Via Amendola 173, 70126 Bari, Italy \\[0.5cm]

and\\[0.5cm]

{\large {\bf A.G. Ushveridze} }\footnote{This work was
partially supported the Deutsche Forschungs Gemainschaft (DFG)
and  by the Lodz University grant no. 457}
\vskip 0.1 cm

Department of Theoretical Physics, University of Lodz,\\
Pomorska 149/153, 90-236 Lodz, Poland\footnote{E-mail
address: alexush@mvii.uni.lodz.pl and alexush@krysia.uni.lodz.pl} \\

\end{center}
\vspace{1 cm}
\begin{abstract}

In this paper the relationship between the problem of
constructing the ground state energy for the quantum
quartic oscillator and the problem of computing mean
eigenvalue of large positively definite random hermitean matrices
is established. This relationship enables one to present several
more or less closed expressions for the oscillator energy.
One of such expressions is given in the form of simple
recurrence relations derived by means of the method of
orthogonal polynomials which is one of the basic tools in
the theory of random matrices.

\end{abstract}

\newpage

\section{Introduction}

In recent years there has been an interest in quantum mechanical
{\it quasi-exactly solvable} problems since they seem related to many
recent developments in theoretical physics \cite{shif,book}.

It was also shown that the Schr\"odinger equation for the energy
eigenvalues for generic, that is  {\it non-quasi-exactly solvable}
potentials, may be approximated by an infinite sequence of potentials
each of which is {\it quasi-exactly solvable}. More specifically,
the solutions of the bound state equation for the quartic anharmonic
oscillator are related to a sequence where the potential is a
even polynomial of degree six. This fact enabled one to formulate
a regular iterative method for constructing the ground
state energy of quartic oscillator as an analitic function
of coupling constant \cite{book}.

In the present paper, it will be shown that  the above mentioned
sequence of approximations is naturally rewritten in terms of the
large $N$ limit for a random matrix\footnote{This is a
particular case of a more general result, according to which,
any quasi-exactly solvable model is equivalent to a certain
random matrix model \cite{cicush}. }. The analysis presented here
still is not a practical approach to the quartic anharmonic
oscillator. However the techniques to analyze large random matrices
had  impressive developments in the past few years, so this approch
seems promising.

After rewriting the problem in terms of random matrix theory, it
appears that the ground state energy of the quartic anharmonic
oscillator is exactly provided by the first few terms in the topological
expansion of a different "potential", in a zero-dimensional
field theory. In other words, the exact sum of Feynman graphs of
a one-dimensional model, which provide ground state energy of the
quartic anharmonic oscillator, are exactly given by the Feynman
graphs of low genus, of a different zero-dimensional model. This
feature is quite unexpected and is the second motivation of the
present paper.

\section{Quasi-exactly solvable sextic oscillator model}

Consider a one-dimensional quantum sextic oscillator model with hamiltonian
\begin{eqnarray}
H_M=-\frac{\partial^2}{\partial x^2}+[b^2-a(4M+3)]x^2+2abx^4+a^2x^6
\label{a.1}
\end{eqnarray}
in which $a>0$ and $b$ are real parameters and $M$ is an
arbitrarily chosen non-negative integer. It is known that
model (\ref{a.1}) is {\it quasi-exactly solvable}. This
means that for any given $M$ the Schr\"odinger equation for
it admits algebraic solutions only for some limited part
of the spectrum. The corresponding (algebraically
calculable) energy levels are given by the formula
\begin{eqnarray}
E_M=(4M+1)b+8a\sum_{i=1}^M \xi_i,
\label{a.2}
\end{eqnarray}
in which $\xi_1,\dots, \xi_M$ are real numbers satisfying
the system of $M$ algebraic equations
\begin{eqnarray}
\sum_{k=1,k\neq i}^M\frac{1}{\xi_i-\xi_k}+\frac{1}{4\xi_i}-b-2a\xi_i=0,
\label{a.3}
\end{eqnarray}
with $i=1,\ldots M$.
It can be shown that system (\ref{a.3}) has $M+1$ solutions
describing energy levels with numbers $0, 2, \ldots, 2M$.
If the number of a level is $2K$ then $K$ of quantities $\xi_i$
are positive and $M-K$ of them are negative. In particular,
for the ground state all the quantities $\xi_i$ should be negative.

\section{Quartic oscillator as a limiting case of
quasi-exactly solvable sextic oscillators}

Let us now demonstrate that if $M$ tends to infinity, the
model (\ref{a.1}) is reduced to the model with quartic anharmonicity.
In order to demonstrate this let us assume that parameters
$a$ and $b$ are functions of $M$. We choose this dependence
in the form
\begin{eqnarray}
a=\frac{1}{2}g^{\frac{2}{3}}f^{-1}(g,M), \qquad
b=g^{\frac{1}{3}}f(g,M),
\label{a.4}
\end{eqnarray}
where $f(g,M)$ is a function of two variables $g$ and $M$
implicitly determined from the equation
\begin{eqnarray}
f^3(g,M)-g^{-\frac{2}{3}}f(g,M)=2(M+\frac{1}{4}).
\label{a.5}
\end{eqnarray}
In this case we will have
\begin{eqnarray}
b^2-a(4M+1)=1, \qquad 2ab=g,
\label{a.6}
\end{eqnarray}
\begin{eqnarray}
H_M=-\frac{\partial^2}{\partial x^2} + (1-2a)x^2 + g x^4 +a^2 x^6
\label{a.7}
\end{eqnarray}
with $a$ being the root of the cubic equation
\begin{eqnarray}
g^2 = 4 a^2 [ 1 + a(1+4 M)]
\label{a.8}
\end{eqnarray}
in the interval $ 0< a< g/2 $.  Since
\begin{eqnarray}
a\approx {1 \over 2} \frac{ g^{2/3}}{(2M)^{1/3}}
 \quad \mbox{for} \quad M\rightarrow \infty, \quad g \; \mbox{fixed}
\label{a.9}
\end{eqnarray}
in the limit $M\rightarrow \infty$ the
model (\ref{a.1}) reduces to the quartic oscillator model
with hamiltonian
\begin{eqnarray}
H=\lim_{M\rightarrow \infty} H_M=-\frac{\partial^2}{\partial x^2}+x^2+gx^4.
\label{a.10}
\end{eqnarray}

Now what about the solutions of the model (\ref{a.10})? It
is quite obvious that at least all its even solutions can be
recovered from formulas (\ref{a.2}) -- (\ref{a.3}) if we
substitute in them the expressions (\ref{a.4}) for $a$ and
$b$ and then take the limit $M\rightarrow \infty$. For the
sake of simplicity we can restrict ourselves by discussing
only the ground state solution for which all the numbers
$\xi_1,\ldots,\xi_M$ should be negative. But
before going over to the limit $M\rightarrow \infty$,
it is desirable to make (for the sake of
further convenience) some appropriate rescaling of
quantities $\xi_1,\dots,\xi_M$, introducing instead of them
the new quantities $\eta_1,\ldots, \eta_M$ by the formulas
\begin{eqnarray}
\xi_i=-g^{-\frac{1}{3}}f^2(g,M)\eta_i, \quad i=1,\ldots, M.
\label{a.11}
\end{eqnarray}
After this the expression for the ground state energy in
the model (\ref{a.10}) takes the form
\begin{eqnarray}
E=\lim_{M\rightarrow \infty} E_M=\lim_{M\rightarrow \infty}
(4M+1)g^{\frac{1}{3}}f(g,M)
\cdot\left[1-S_M(\bar\eta_1,\ldots,\bar\eta_M) \right],
\label{a.12}
\end{eqnarray}
where
\begin{eqnarray}
S_M(\eta_1,\ldots,\eta_M)=\sum_{i=1}^M\frac{\eta_i}{M+1/4}
\label{a.13}
\end{eqnarray}
and $\bar\eta_1,\dots, \bar\eta_M$ are
positive solutions of the system
\begin{eqnarray}
\sum_{k=1,k\neq i}^M\frac{1}{\eta_i-\eta_k}+\frac{1}{4\eta_i}
+ f^3(g,M)(1-\eta_i)=0,
\label{a.14}
\end{eqnarray}
with $i=1,\ldots M$.

Taking into account the facts that the large $M$
asymptotics of function $f(g,M)$ is
\begin{eqnarray}
f(g,M)=2^{\frac{1}{3}}(M+1/4)^{\frac{1}{3}}\left\{
1+\frac{1}{3[2g(M+1/4)]^{\frac{2}{3}}}+\ldots
\right\},
\label{a.15}
\end{eqnarray}
and the ground state energy $E$ of the quartic anharmonic
oscillator is a finite number,
we can conclude that for large values of $M$ the quantity
$S_M(\bar\eta_1,\ldots,\bar\eta_M)$ should behave as
\begin{eqnarray}
S_M(\bar\eta_1,\ldots,\bar\eta_M)=
1-\frac{E}{4(2g)^\frac{1}{3}}\cdot\frac{1}{M^\frac{4}{3}}+\ldots
\label{a.16}
\end{eqnarray}
In other words, if $S_M(\bar\eta_1,\ldots,\bar\eta_M)$
is already known and we would like to find
the ground state energy of the quartic anharmonic
oscillator, it is sufficient to find the first
non-vanishing correction to unity in the expansion of
$S_M(\bar\eta_1,\ldots,\bar\eta_M)$
in negative powers of $M$. All higher corrections will be
irrelevant to us.

\section{Integral representation for the ground state
energy in quartic oscillator model}

The result of the previous section
enables one to replace $S_M(\bar\eta_1,\ldots,\bar\eta_M)$
in formula (\ref{a.12})
by another quantity which we denote by $S_M$ and define as
\begin{eqnarray}
S_M=  \sum_{i=1}^M \frac{<\eta_i>}{M+1/4} =
\frac{\int_0^\infty D\eta S_M(\eta_1,\ldots,\eta_M) \exp
[-V(\eta_1,\ldots,\eta_M)]}{\int_0^\infty D\eta
\exp[- V(\eta_1,\ldots,\eta_M)]}
\label{a.17}
\end{eqnarray}
where
\begin{eqnarray}
D\eta=\prod_{i=1}^M d\eta_i
\label{a.18}
\end{eqnarray}
and
\begin{eqnarray}
V(\eta_1,\ldots,\eta_M)=-\sum_{i\neq k}^M\ln|\eta_i-\eta_k|-
\frac{1}{2}\sum_{i=1}^M \ln|\eta_i|+f^3(g,M)\sum_{i=1}^M(\eta_i-1)^2
\label{a.19}
\end{eqnarray}
The reason for such a replacement can be explaned as follows.
It is clear that the exponent (\ref{a.19}) has a minimum
which is saturated by numbers $\bar\eta$ satisfying the
system of equations (\ref{a.14}). As follows from these
equations, all the numbers $\bar\eta$ should
be of order of unity, and therefore,
the most relevant values of exponent (\ref{a.19})
should be of order $M^2$.  This means that
for large values of $M$ the multiple integrals
in (\ref{a.17}) can be evaluated by the
saddle point method and the correction to the zero order
approximation should be of order $M^{-2}$.
Now note that the zero order approximation for (\ref{a.17})
is nothing else than $S_M(\bar\eta_1,\ldots,\bar\eta_M)$. The fact that the
difference between $S_M(\bar\eta_1,\ldots,\bar\eta_M)$
and $S_M$ is of order $M^{-2}$
(i.e. less than $M^{-\frac{4}{3}}$ ) makes it possible to
replace $S_M(\bar\eta_1,\ldots,\bar\eta_M)$ by $S_M$ in the
expression for $E$.
This enables one to write the following simple formula for
the ground state energy in the quartic oscillator model (\ref{a.10}):
\begin{eqnarray}
E=\lim_{M\rightarrow \infty}
(4M+1)g^{\frac{1}{3}}f(g,M)
\cdot\left[1-S_M \right],
\label{a.20}
\end{eqnarray}
where $S_M$ is defined by formulas  (\ref{a.17}) -- (\ref{a.19}).

\section{Random matrix model}

In this section we recall that the expression (\ref{a.17}) for $S_M$
naturally arises in the theory of random matrix models.

Consider a zero-dimensional "field-theoretical" model in which the role
of a field is played by the random $M\times M$ positively
definite hermitean
matrix $A$. Let us choose the action of the model in the form
\begin{eqnarray}
W_M[A]=\mbox{Tr}[-\frac{1}{2}\ln A + f^3(g,M)(A-1)^2]
\label{a.21}
\end{eqnarray}
which is obviously invariant under unitary transformations
$A\rightarrow U A U^{-1}$.
The simplest invariant quantities in this model are
the eigenvalues of the matrix $A$. Consider the problem of
computing  $\bar A = \langle \mbox{Tr} A \rangle/(M+1/4)$.
The result can be represented in
the form of a multiple integral
\begin{eqnarray}
\bar A
=\frac{1}{M+1/4}
\frac{\int DA \mbox{Tr} A\exp\{-W_M[A]\}}{\int DA \exp\{-W_M[A]\}}
\label{a.22}
\end{eqnarray}
which can be simplified if one uses the well-known
Faddeev-Popov method for extracting from (\ref{a.22}) the volume of the
unitary group. Using formulas
\begin{eqnarray}
A=U \cdot \mbox{diag}\{\eta_1,\ldots,\eta_M\}\cdot U^{-1}
\label{a.23}
\end{eqnarray}
and
\begin{eqnarray}
DA=\prod_{i=1}^M d\eta_i \prod_{i<k}(\eta_i-\eta_k)^2 DU
\label{a.24}
\end{eqnarray}
in which $\eta_i$ are the eigenvalues of $A$ and $DU$
denotes the measure on the unitary group, we can easily
reduce the expression (\ref{a.22}) to the form
\begin{eqnarray}
\bar A
=\frac{\int_0^\infty D\eta S_M(\eta_1,\ldots,\eta_M) \exp
[-V(\eta_1,\ldots,\eta_M)]}{\int_0^\infty D\eta
\exp[- V(\eta_1,\ldots,\eta_M)]}
\label{a.25}
\end{eqnarray}
in which $D\eta$, $S_M(\eta_1,\ldots,\eta_M)$ and $V(\eta_1,\ldots,\eta_M)$
are given by formulas (\ref{a.18}), (\ref{a.13}) and
(\ref{a.19}), respectively. We see that formulas
(\ref{a.25}) and (\ref{a.17}) exactly coincide!
This means that the problem of
finding the ground state energy in the quartic oscillator
model (\ref{a.10}) is equivalent to the problem of finding
the mean eigenvalue of the random matrix $A$ in the theory with
the action (\ref{a.21}).

\section{Orthogonal polynomials and recurrence relations}

The equivalence of models (\ref{a.10}) and (\ref{a.21})
enables one to simplify the
expression (\ref{a.20}) by using the method of orthogonal polynomials
which is one of the basic tools in the theory of random
matrix models \cite{bessis}.

The idea of this method can be formulated as follows.
Let us rewrite the expression (\ref{a.17}) in the form
\begin{eqnarray}
S_M=\frac{\int_0^\infty \prod_{i=1}^M d\eta_i S_M(\eta_1,\ldots,\eta_M)
\prod_{i<k}^M (\eta_i-\eta_k)^2 \exp
[-V_0(\eta_1,\ldots,\eta_M)]}
{\int_0^\infty \prod_{i=1}^M d\eta_i
\prod_{i<k}^M (\eta_i-\eta_k)^2\exp
[-V_0(\eta_1,\ldots,\eta_M)]}
\label{a.26}
\end{eqnarray}
with
\begin{eqnarray}
V_0(\eta_1,\ldots,\eta_M)=
\frac{-1}{2}\sum_{i=1}^M \ln|\eta_i|+f^3(g,M)\sum_{i=1}^M(\eta_i-1)^2
\label{a.27}
\end{eqnarray}
and note that
\begin{eqnarray}
\prod_{i<k}^M (\eta_i-\eta_k)^2=[\det || \eta_i^{k-1}||_{i,k=1,\ldots,M}]^2=
[\det ||P_{k-1}(\eta_i)||_{i,k=1,\ldots,M}]^2
\label{a.28}
\end{eqnarray}
where $P_n(\eta)$ denote arbitrary linearly independent polynomials
of degrees $n$ normalized as $P_n(\eta)=\eta^n+\ldots$.
Because of the arbitrariness of these polynomials we can
consider them (without any loss of generality) as
polynomials orthogonal with the weight $|\eta|^{1/2}\exp[-f^3(g,M)(\eta-1)^2]$,
i.e. requiring the condition
\begin{eqnarray}
\int_{0}^\infty d\eta P_n(\eta)P_m(\eta) |\eta|^{1/2}\exp[-f^3(g,M)(\eta-1)^2]
=h_n \, \delta_{m n}
\label{a.29}
\end{eqnarray}
This immediately leads us to the formula
\begin{eqnarray}
S_M=\frac{1}{M+1/4}\sum_{m=0}^{M-1}\frac{
\int_{0}^\infty d\eta \eta P_m^2(\eta) |\eta|^{1/2}\exp[-f^3(g,M)(\eta-1)^2]}
{ h_m}
\label{a.30}
\end{eqnarray}

By using well known methods, \cite{gross}, recurrence relations are
derived for the orthogonal polynomials, which allow the evaluation
of the expression (\ref{a.30}). The polynomials are given by a three terms
equation

\begin{eqnarray}
\eta P_n(\eta)=P_{n+1}(\eta) + A_n(\eta)P_n(\eta) +R_n(\eta)P_{n-1}(\eta)
\label{a.31}
\end{eqnarray}

with $ R_n(\eta) = \frac{h_n}{h_{n-1}} $. It follows that

\begin{eqnarray}
S_M= \frac{1}{M+1/4} \sum_{k=0}^{M-1} A_k
\label{a.32}
\end{eqnarray}

It is also customary to introduce the states $ |n> $

\begin{eqnarray}
<\eta | n > = P_n(\eta), \quad \mbox{where} \quad <m |n > = \delta_{m n}
\label{a.33}
\end{eqnarray}

One obtains \cite{morris} a local string equation
\begin{eqnarray}
< n| \eta W'(\eta) | n> = 2n +1
\label{a.34}
\end{eqnarray}
that is
\begin{eqnarray}
2 f^3 (g,M) [ A_n^2 - A_n + R_{n+1} + R_n ] = 2n + \frac{3}{2}
\label{a.35}
\end{eqnarray}
and a non-local equation

\begin{eqnarray}
\sum_{k=0}^{n} A_k &=& \sqrt{R_{n+1} } < n+1 |  \eta W'(\eta) | n> =
\nonumber \\
&=&  R_{n+1} \, 2 \, f^3 (g,M) (A_{n+1} + A_n -1)
\label{a.36}
\end{eqnarray}

which leads to a further expression for $ S_M $
\begin{eqnarray}
S_M= \frac{ 2 f^3 (g,M)}{ M+1/4} R_M [A_M +A_{M-1} -1 ]
\label{a.37}
\end{eqnarray}

If the coefficients $ A_k, \, R_k $ are known up to $ k=n$, eq. (\ref{a.35})
allows the determination of $ R_{n+1} $ , next $A_{n+1}$ may be evaluated
by eq. (\ref{a.36}).  The lowest values are

\begin{eqnarray}
A_0 =\frac{\mu_1}{\mu_0}  &,&  R_0=0  \nonumber \\
A_1 =1-\frac{\mu_1}{\mu_0}+\frac{1}{2f^3(g,M) (\frac{\mu_2}{\mu_1}-
\frac{\mu_0}{\mu_1})} &,&  R_1= \frac{\mu_2}{\mu_0}-A_0^2
\label{a.38}
\end{eqnarray}
where $ \mu_n $ are the moments
\begin{eqnarray}
\mu_n (g,M)= \int_{0}^\infty dx \, x^{n+1/2} e^{-f^3(g,M) \, (x-1)^2}
\label{a.39}
\end{eqnarray}
Since $ S_M $ in eq.(\ref{a.37}) determines the ground energy $E(g)$
of the quartic anharmonic oscillator (\ref{a.20}) for large values of $M$,
the above iterations do not seem very useful. It might seem more promising
to derive a recurrence relation immediately for the quantities
\begin{eqnarray}
q_n=\sum_{k=0}^{n-1} A_k
\label{a.40}
\end{eqnarray}
One easily obtains:
\begin{eqnarray}
(q_{n+1}-q_n &)
 &(q_{n+1}-q_n-1) + \frac{1}{2f^3(g,M)}
 [ \frac{q_{n+1}}{q_{n+2}-q_n-1} \nonumber \\
&+& \frac{q_n}{q_{n+1}-q_{n-1}-1} -2n-3/2 ] =0
\label{a.41}
\end{eqnarray}
where $q_{n+2}$ is determined in terms of $q_{n+1}, q_n, q_{n-1}$. The lowest
values are $q_0 =0, q_1= A_0, q_2= A_0+A_1 $

Thus the final formula for the ground state energy of the
quartic oscillator model (\ref{a.10}) has the form
\begin{eqnarray}
E=\lim_{M\rightarrow \infty} E_M=\lim_{M\rightarrow \infty}
4(2M^4 g)^{\frac{1}{3}}\cdot\left[1-\frac{q_M}{M+1/4}\right].
\label{a.42}
\end{eqnarray}
where $q_M$ can be found from the recurrence relation (\ref{a.41}).

\section{Conclusion}

The equations (\ref{a.41}) and (\ref{a.42}) obtained
in the above Section are {\it exact} and, in principle,
provide an {\it exact} evaluation of $E(g)$ (i.e. with an
arbitrarily high accuracy). It seems, however, that these
equations have yet a purely theoretical significance,
because the practical use of these formulas encounters
serious difficulties.

The first difficulty follows from the fact that the
deviation of the exact value of $E$ from its approximations $E_M$ obtained for
finite $M$'s is of order $M^{-2/3}$. Indeed, remember, that
the energy $E$ appeared as a coefficient in front of
$M^{-4/3}$ in the expansion of $S_M(\bar\eta_1,\ldots,\bar\eta_M)$
(see formula (\ref{a.16}) ), and we replaced
$S_M(\bar\eta_1,\ldots,\bar\eta_M)$ by $S_M$ after
neglecting corrections of order $M^{-2}$ (see Section 4).
This means that the sequence of $E_M$ converges to $E$ very
slowly. For example, if we want to compute $E$ with 1
percent accuracy, we should take $M=1000$, and this means
that we need 1000 steps of the recurrence procedure (\ref{a.41}).
The recurrence relations (\ref{a.41}) are, however, rather unstable,
and the iteration of $q_n$ from
low to high values of $n$ does not seem practical, because
of the rapid accumulation of errors.

The second difficulty lies in the fact that the direct analysis of the
asymptotic behaviour of the difference equation (\ref{a.41}) is
non trivial. It is rather easy to obtain the leading
behaviour of numbers $q_n$ for large $n$, considering $q_n$
as an analytic function of $n$, $q_n=q(n)$ and solving the
associated differential equation for $q(n)$. The result, $q_n
\sim n+1/4$, does not contradict formula (\ref{a.42}).
However, a naive attempt of finding in such a way the non-leading
behaviour of $q(n)$, which is relevant for eq.(\ref{a.42}), leads
to {\it a priori} false result.

It seems that both above mentioned difficulties are closely
related to each other. It is possible that function
$q(n)$ is continuous, but not differentiable. In other
words, the first (or, perhaps, the second) derivative of
$q(n)$ might be a random function of $n$. In this case, the
approximation of the recurrence relations (\ref{a.41}) by a
differential equation is, generally, rather rough, and
cannot be used for studying any subtle properties of function $q_n$.
If the behaviour of the differences $q_n-q_{n-1}$ (the
derivatives of $q(n)$) is really stochastic, then this fact would
easily explain the rapid accumulation of errors in
numerical computations of the ground state energy of quartic
oscillator by formulas (\ref{a.41}) and (\ref{a.42}).

In conclusion of this Section we demonstrate that the
problem of determination of a non-leading behaviour of
numbers $q_n$ is related to the problem of finding the form
of the second local string equation for the recurrence relations.
In order to show this, let us first provide an approximate evaluation
of $S_M$ for large $M$ by means of the saddle point method.
At leading order, for large $M$, one may replace the discrete eigenvalues
$ \eta_i$ with the continuous function $ \eta(i/M) =\eta_i$ and the algebraic
system (\ref{a.14}) with the singular integral equation
\begin{eqnarray}
\intcauchy_0^1 dy \frac{1}{\eta(x)-\eta(y)} + \frac{1}{4M\eta(x)} +
\frac{f^3(g,M)}{M} \left[ 1-\eta(x) \right] =0
\label{a.43}
\end{eqnarray}
which, after the introduction of the eigenvalue density $ \rho(\eta)=
\frac{dx}{d\eta} $, becomes
\begin{eqnarray}
\intcauchy_A^B d\mu \frac{\rho(\mu)}{\lambda -\mu} =
\frac{f^3(g,M)}{M} \left[\lambda-1 \right] -\frac{1}{4M\lambda}
\label{a.44}
\end{eqnarray}
This is the saddle-point approximation for the large-$M$ limit of the matrix
model (\ref{a.21}). Its solution with $0<A<B$ was described in \cite{noi},
together with the evaluation of expectation values. One obtains, for
large $M$
\begin{eqnarray}
 \sum_{i=1}^M <\eta_i> \sim M\int_A^B \mu \rho(\mu) d\mu =
(B-A)^2 (A+B-1) \frac{f^3(g,M)}{8}
\label{a.45}
\end{eqnarray}
where the extrema of the support are solution of the algebraic equations
\begin{eqnarray}
2(A+B)^2+(B-A)^2 -4(A+B)&=&\frac{4M}{f^3(g,M)}(2+\frac{1}{2M}) \nonumber \\
 2\sqrt{AB} (A+B-2) f^3(g,M)&=&1
\label{a.46}
\end{eqnarray}
 From the system (\ref{a.46}) one easily evaluates the large $M$ expansions
\begin{eqnarray}
A=\frac{a_1}{M^{2/3}} +\frac{a_2}{M^{4/3}}+ ..\nonumber \\
B=2 + \frac{b_1}{M^{2/3}}+\frac{b_2}{M^{4/3}}+ ..
\label{a.47}
\end{eqnarray}
To estimate the present saddle-point approximation, we quote the first
coefficients in the limit of large values of $g$
\begin{eqnarray}
a_1=\frac{1}{2^{5/3}} [1+O(g^{-2/3})]  &,& b_1= \frac{-1}{2 (2g)^{-2/3}}
\nonumber \\
a_2=\frac{1}{32} \, \frac{1}{2^{1/3}}[1+O(g^{-2/3})] &,&
 b_2= \frac{-3}{64}\,\frac{1}{2^{1/3}} [1+O(g^{-2/3})]
\label{a.48}
\end{eqnarray}
Then
\begin{eqnarray}
S_M=  \sum_{i=1}^M \frac{<\eta_i>}{M+1/4}
& \sim &1-\frac{1}{M^{4/3}}[\frac{3}{16}\frac{1}{2^{1/3}}+O(g^{-2/3})]
\nonumber \\
E(g) &\sim & g^{1/3}\; \frac{3}{4} [1+O(g^{-2/3})]
\label{a.49}
\end{eqnarray}
The comparison of the approximate result
(\ref{a.49}) with the known first term of the strong coupling
expansion \cite{montroll}
\begin{eqnarray}
E(g) = 2^{2/3} g^{1/3} [0.667 +O(g^{-2/3})]
\label{a.54}
\end{eqnarray}
shows that the fractional powers of the
coupling constant $g$ are correctly reproduced, but
the relative error for the first coefficient in the strong coupling
limit of the ground state energy is about $32 \%$.
This fact is not surprizing because we started with a
continuous version of the saddle point equation, which is
only an approximation to the true discrete one.

Let us now consider the equivalent approximation by recurrence relation.
If we neglect the contributions in integration by parts at $\eta=0$,
discussed in \cite{morris}, a second local string equation is obtained
\cite{gross} :
\begin{eqnarray}
<n |W'(\eta)|n>=0
\label{a.50}
\end{eqnarray}
and evaluating it in the continuum approximation ( that is $ A_n =A(
\frac{n}{M}) \, , \, R_n=R(\frac{n}{M}) )$ and neglecting the derivatives,
it reads
\begin{eqnarray}
4f^3(g,M) [A(x)-1] = \frac{1}{\sqrt{A^2(x)-4R(x)}}
\label{a.51}
\end{eqnarray}
The same approximation on the first local string eq.(\ref{a.35}) is
\begin{eqnarray}
\frac{2f^3(g,M)}{M} [A^2(x)-A(x)+2R(x)] =2x+\frac{3}{2M}
\label{a.52}
\end{eqnarray}
This algebraic system, for large values $n \sim M $ that is $x \sim 1$
reproduces the system (\ref{a.46}) after the usual correspondence \cite{nota}
\begin{eqnarray}
R(1)= \frac{(B-A)^2}{16} \nonumber \\
A(1)= \frac{(B+A)}{2}
\label{a.53}
\end{eqnarray}
{}From the above exercise we learn that to improve on the approximate
evaluation (\ref{a.49}) by use of the recurrence relations, it is likely
to need an improved evaluation of the second string equation, by including
the derivatives and the boundary contributions at $\eta =0$.
We intend to consider this question in fortcoming publications.

\section{Acknowledgements}

One of us (AGU) is grateful to Prof.Dieter Mayer for interesting discussions.
He also thanks the staff of Theoretical Physics Department
of the University of Parma for kind hospitality.


\begin{thebibliography}{99}
\bibitem{nota} The non-leading difference of the term $(2+\frac{1}{2M})$
on the right side of eq.(\ref{a.46}) versus the term $(2+\frac{3}{2M})$
on the right side of eq.(\ref{a.52})
is irrelevant in this context. Indeed both the approximate methods in
this Section are equivalent at leading order for large $M$ and fail
to reproduce the relevant non leading behaviour.
\end{thebibliography}
\end{document}